\documentclass{article}
\usepackage{spconf,amsmath,graphicx, amssymb, booktabs}

\usepackage{cite}
\usepackage{bm}
\usepackage{multirow}

\usepackage{mathrsfs}
\usepackage{url}

\title{ATTENTION-BASED SCALING ADAPTATION FOR TARGET SPEECH EXTRACTION}
\name{Jiangyu Han$^{1, 2}$, Wei Rao$^{2}$, Yanhua Long$^{1}$, Jiaen Liang$^3$\thanks{Work was done when Jiangyu Han was an intern at Tencent Ethereal Audio Lab. Yanhua Long is the corresponding author. The work is supported by the National Natural Science Foundation of China (Grant No.62071302 and
No.61701306).}}
\address{
  $^1$Shanghai Normal University, Shanghai, China\\
  $^2$Tencent Ethereal Audio Lab, Tencent Corporation, Shenzhen, China\\
  $^3$Unisound AI Technology Co., Ltd., Beijing, China}
\begin{document}
\ninept
\maketitle
\begin{abstract}
The target speech extraction has attracted widespread attention
in recent years. In this work, we focus on investigating the dynamic interaction
between different mixtures and the target speaker to exploit the discriminative
target speaker clues. We propose a special attention mechanism without introducing
any additional parameters in a scaling adaptation
layer to better adapt the network towards extracting the target speech.
Furthermore, by introducing a mixture embedding matrix pooling method,
our proposed attention-based scaling adaptation (ASA) can exploit the
target speaker clues in a more efficient way.
Experimental results on the spatialized reverberant WSJ0 2-mix dataset demonstrate that the proposed method can improve the performance of the target speech extraction effectively.
Furthermore, we find that under the same network configurations,
the  ASA in a single-channel condition
can achieve competitive performance gains as that achieved from
two-channel mixtures with inter-microphone phase difference (IPD) features.
\end{abstract}
\section{Introduction}
\label{sec:intro}

Speech separation aims to separate each source
signal from the mixed speech. Many works have been proposed
to perform the separation either in the time-frequency domain or the purely time-domain,
such as deep clustering \cite{dpcl}, deep attractor network \cite{dan}, permutation invariant training (PIT) \cite{pit} and the recent convolutional time-domain audio separation network (Conv-TasNet) \cite{tasnet}, etc. Our work aims to generalize the time-domain Conv-TasNet
to the target speech extraction (TSE) tasks, because of its higher performances
over most time-frequency domain methods \cite{tasnet_relate1, tasnet_relate2, tasnet_relate3, tasnet_relate4, tasnet_relate5, tasnet_relate6}.
Compared with the pure speech separation techniques, most TSE approaches require additional target speaker clues to guide the separation network to extract the speech of that speaker.

Many works have been proposed for the TSE target speaker adaptation, such as the
SpeakerBeam-FE \cite{sbfe}, SpEx \cite{spex}, and time-domain SpeakerBeam
(TD-SpeakerBeam) \cite{td_speakerbeam} employed a factorized
adaptation layer, a concatenating adaptation layer, and a scaling
adaptation layer to adjust the internal behavior of the separation network respectively.
Although these adaptation methods showed promising results,
their exploitations of the target speaker clues are somehow monotonous.
Specifically, they average the speaker vectors over an adaptation
or enrollment utterance to get a vector that provides
a global bias for the target speaker, then, repeat the same
vector for each frame adaptation of the mixed speech
to drive the network towards extracting the target speech,
no matter how the acoustic characteristics
of the frame are related to the target speaker identity.
Therefore, we think that a more reasonable target speaker bias
should be able to adjust itself dynamically according to the
acoustic interaction between the mixture and the target speaker.

In \cite{ms}, an attention mechanism between the adaptation utterances
and the mixed speech was proposed to dynamically compute a
frame-wise speaker bias which is a weighted sum of target speaker embeddings.
However, it was designed for the time-frequency
domain TSE tasks, it is tricky to generalize such attention
to the time-domain TSE tasks. Because the frames of the encoded
representation for a waveform in the time-domain (e.g. 3199)
is usually much larger than that in the time-frequency domain (e.g. 251).
Such attention performed on these large frames may result in a very
sparse distribution of their softmax probabilities, in which
only several frames have big probabilities, while others are close
to zero. In this case, the weighted sum of target speaker vectors
can no longer provide a sufficient discriminative bias for extracting
the target speech. Although the authors in \cite{tencent} applied
the same attention mechanism to the time-domain tasks,
they only presented the results of the pre-trained speaker embedder.

In this study, we design a novel attention mechanism to exploit the interaction
between the adaptation utterance and different mixture speech for the time-domain TSE tasks.
Different from \cite{ms}, our attention is performed on the global bias vector
of the target speaker and the embedding matrix of the mixed speech,
because we want to adjust the target speaker bias dynamically according to
the acoustic interaction between the mixture and target speaker
at different time intervals. Rather than learning the context-dependent
information in \cite{ms}, we pay more attention to exploit the dynamic
target speaker identity clues between different mixtures and the target
speaker adaptation utterance. Furthermore,
as our attention is directly performed between a vector and a matrix,
unlike the traditional self-attention \cite{self-att} with linear projection weights,
there are no additional network parameters is introduced in our ASA,
and it has a less computational cost and lower memory requirements
than the attention performed on two matrices.
All experiments are performed on the publicly available
spatialized reverberant WSJ0 2-mix dataset.
Results show that the proposed method improves the target speech
extraction performance effectively, and the single-channel
performances are comparable to that in multi-channel
condition with IPD features \cite{ipd}.

The rest of this paper is organized as follows. In Section \ref{sec:td-speakerbeam}, we introduce arcthitecture of our baseline. In Section \ref{sec:proposed}, we describe the principle of our proposed ASA. 
The experiments and results are presented in Section \ref{sec:exp_res}, and conclude in Section \ref{sec:conclusion}.

\section{Time-domain speakerbeam}
\label{sec:td-speakerbeam}

\begin{figure}[t]
  \centering
  \includegraphics[width=8.5cm]{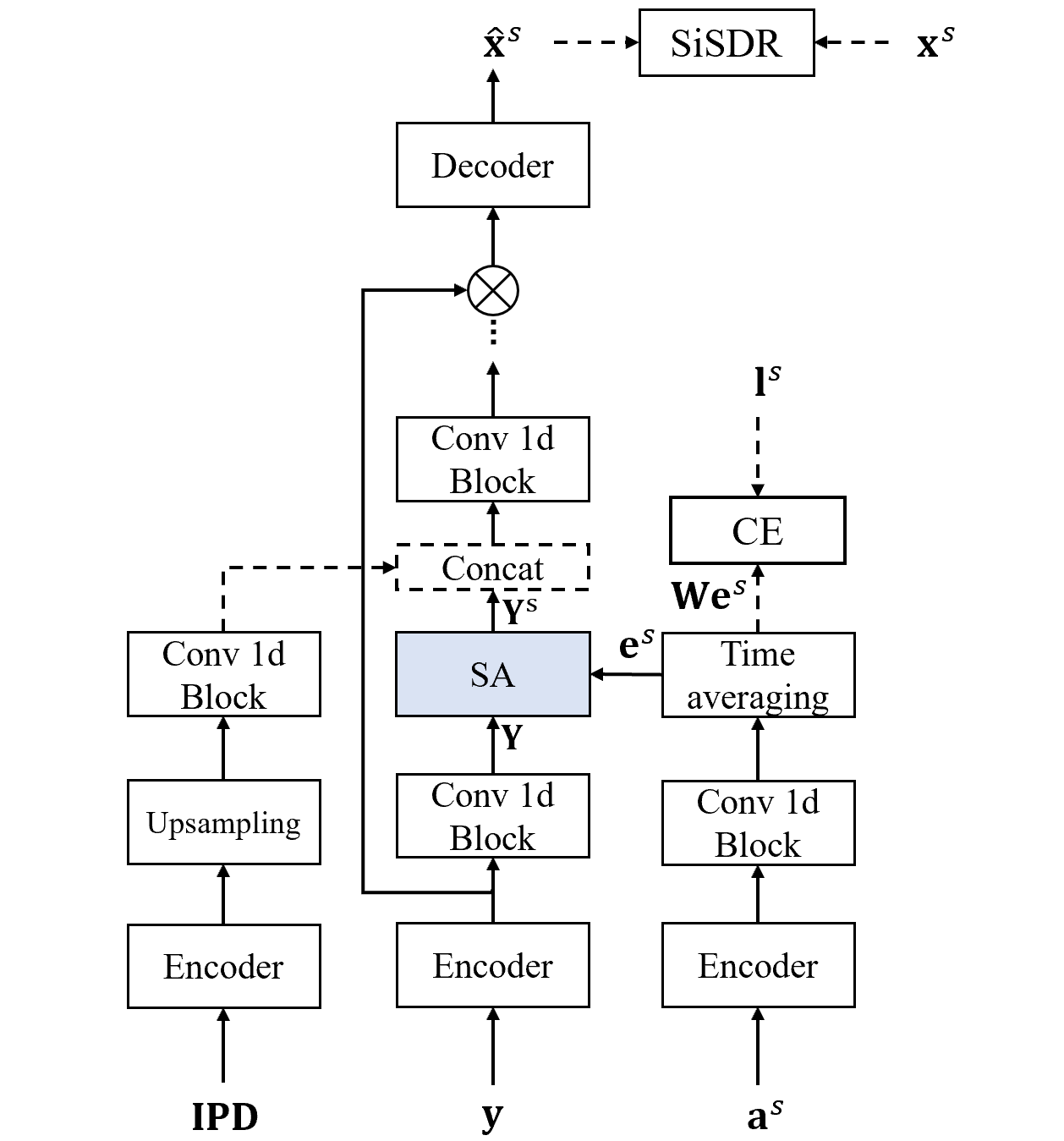}
  \caption{The block diagram of the TD-SpeakerBeam with IPD. ``SA'' represents scalling adaptation.}
  \label{fig:td_speakerbeam}
\end{figure}

TD-SpeakerBeam is a very effective
target speech extraction approach that has been
recently proposed in \cite{td_speakerbeam}.
The structure of TD-SpeakerBeam with IPD concatenation block is shown
in Fig.\ref{fig:td_speakerbeam}.
It contains a 1d convolution encoder,  several convolution
blocks, and a 1d deconvolution decoder.
The $\mathbf{y}$, $\mathbf{\hat{x}}^s$ and $\mathbf{a}^s$
are the mixture waveform, the extracted target speech waveform, and the adaptation
utterance of the target speaker respectively.

The whole TD-SpeakerBeam
network follows a similar configuration as Conv-TasNet\cite{tasnet}
except for inserting a scaling adaptation (SA) layer \cite{adap}
between the first and second convolution blocks to drive the network
towards extracting the target speech.
The output of the SA layer $\mathbf{Y}^{\rm s} \in \mathbb{R}^{N \times T}$
is obtained by simply performing an element-wise multiplication
between the repeated target speaker embedding vectors $[\mathbf{e}^s, \mathbf{e}^s, ..., \mathbf{e}^s]_{N \times T}$ and the input $\mathbf{Y}$ as follows,

\begin{equation}
\mathbf{Y}^{\rm s} = [\mathbf{e}^s, \mathbf{e}^s, ..., \mathbf{e}^s]_{N \times T} \odot \mathbf{Y}
\end{equation}
where the $\mathbf{e}^s \in \mathbb{R}^{N \times 1}$ is computed from a time-domain convolutional
auxiliary network as shown in the right part of Fig.\ref{fig:td_speakerbeam},
and the $N$ is the dimension of the embedding vectors, $T$ is the number of
frames of the convolutional output.
Furthermore, as shown in the left part of Fig.\ref{fig:td_speakerbeam}, TD-SpeakerBeam
can also be extended to the multi-channel TSE task by combining the IPD
features (processed with a 1d convolutional layer, upsampling, and
a convolution block) after the SA layer.

The whole network  of TD-SpeakerBeam is trained jointly in an end-to-end multi-task way. The multi-task loss combines the scale-invariant signal-to-distortion ratio (SiSDR) \cite{sisdr} as
the signal reconstruction loss and cross-entropy as the speaker identification loss. The overall loss function is defined as,
\begin{equation}\small
L(\Theta|\mathbf{y}, \mathbf{a}^s,\mathbf{x}^s,\mathbf{l}^s) = {\rm -SiSDR}(\mathbf{x}^s, \mathbf{\hat{x}}^s) + \alpha{\rm CE}(\mathbf{l}^s, \sigma(\mathbf{We}^s))
\label{eq:mtl}
\end{equation}
where $\Theta$ represents the model parameters, $\mathbf{x}^s$ is the target speech, $\mathbf{l}^s$ is a one-hot vector representing the target speaker identity, $\alpha$ is a scaling parameter, $\mathbf{W}$ is a weight matrix and $\sigma(\cdot)$ is softmax operation.
More details can be found in \cite{td_speakerbeam}.

\label{sec:prop}
\begin{figure}[t]
  \centering
  \includegraphics[width=9cm]{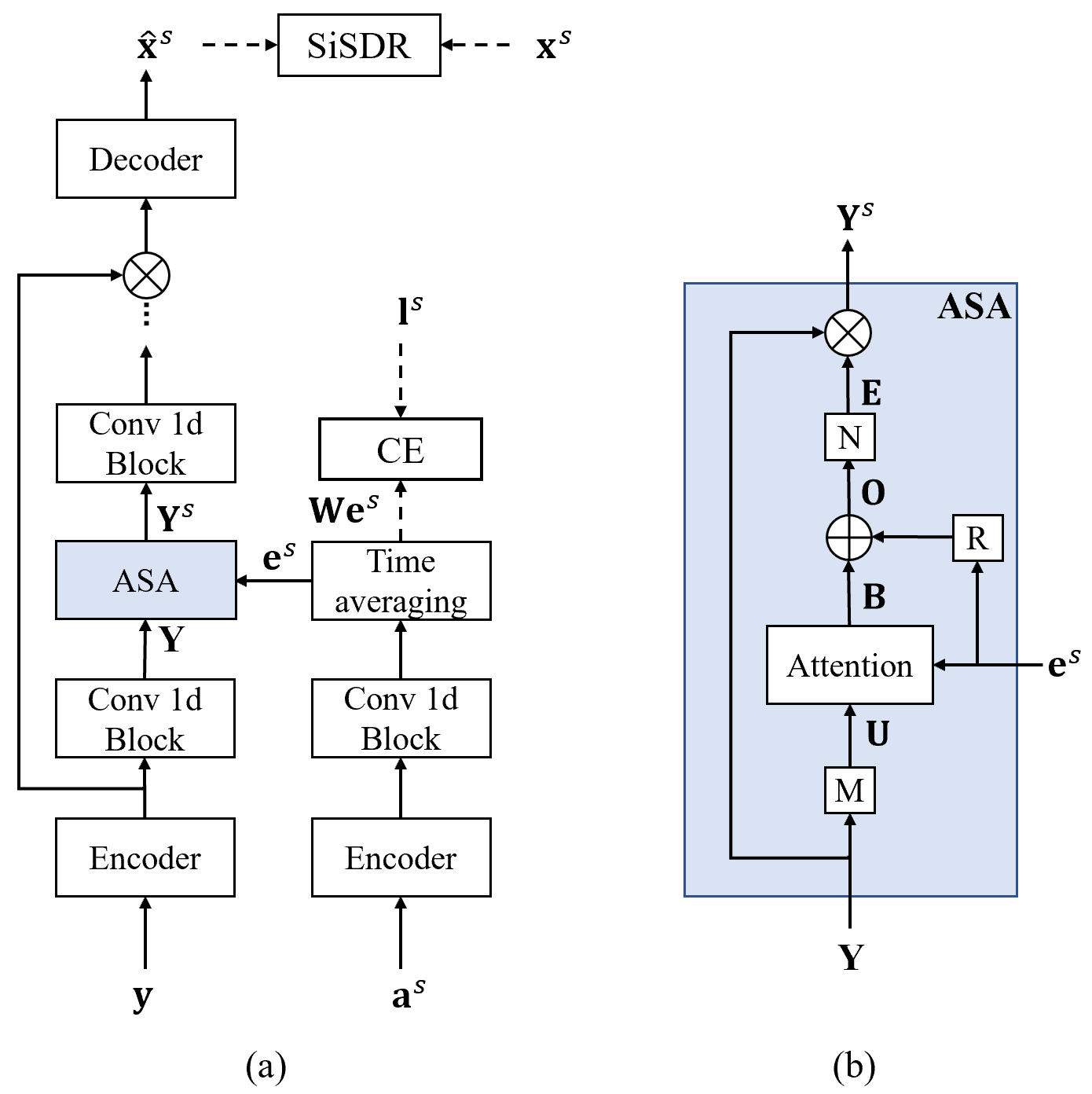}
  \caption{The block diagram of (a) proposed method based on TD-SpeakerBeam and (b) the structure of ASA.
``M'' represents mean pooling, ``R'' represents repeating, and ``N'' represents nearest upsampling.}
  \label{fig:asa}
\end{figure}

\section{Proposed method}
\label{sec:proposed}

\subsection{Overview}
\label{ssec:overiew}

The whole structure of the proposed method is shown in Fig.\ref{fig:asa} (a).
All network architectures are the same as  Fig.\ref{fig:td_speakerbeam}
except for replacing the SA layer with our proposed attention-based scaling
adaptation (ASA).  As shown in Fig.\ref{fig:asa} (b), the ASA layer accepts the mixture
embedding matrix $\mathbf{Y} \in \mathbb{R}^{N \times T}$
and the target speaker embedding vector $\mathbf{e}^s$ as inputs, and outputs
the scaled mixture embedding matrix $\mathbf{Y}^s \in \mathbb{R}^{N \times T}$
whose each value is weighted by the dynamic target speaker bias $\mathbf{E}$.
$N$ is the output dimension of the convolutional network, and $T$ is the number of
frames of the convolutional output.

As our operations are performed in the time-domain, in order to get a high time-domain resolution,
the convolutional kernel used to encode the waveform usually has a small size, which means the
number frames of the convolution output $T$ has a large size.
For example, assuming the input waveform has $32000$ samples, the 1d convolutional kernel size is $20$, stride is $10$ and padding is 0. Then the number of frames of the convolutional output will be $3199$.
When performing the softmax over these frames, the obtained probability vector usually has
a very sparse distribution. Such a sparse distribution will prevent the conventional attention mechanism from being used to effectively exploit the relationship between the mixture and the target speaker for the time-domain tasks.

\subsection{Attention-based scaling adaptation}
\label{sec:asa}

Considering that the speaker characteristics of several
consecutive frames are almost unchanged, we treat each
non-overlapped consecutive $M$ frames as an integral part,
by averaging each $M$ column vectors of the mixture
embedding matrix $\mathbf{Y}$ to generate a speaker-dependent
vector $\mathbf{u}_t$.
After such a mean pooling operation, we can obtain a
speaker-dependent matrix $\mathbf{U}$, i.e.,

\begin{equation}	
\begin{split}
\mathbf{u}_t &= \frac{1}{M}\sum_{j=1}^M{\mathbf{y}_{j+(t-1)M}}\\
\mathbf{U} &= [\mathbf{u}_1, \mathbf{u}_2,..., \mathbf{u}_{T_m}]
\end{split}
\end{equation}
where $\mathbf{U} \in \mathbb{R}^{N \times T_m}$, $T_m$ equals to $T / M$,
$\mathbf{u}_t \in \mathbb{R}^{N \times 1}$ is the $t$-th column vector of $\mathbf{U}$,
and $\mathbf{y}_j \in \mathbb{R}^{N \times 1}$
is the $j$-th column vector of $\mathbf{Y}$.
By doing so, the new mixture embedding matrix
$\mathbf{U}$ not only has a much smaller size than $\mathbf{Y}$,
but also has a relatively compact acoustic representation
to reflect the speaker dependent information, each $\mathbf{u}_i$
summaries the speaker information over several frames and represents
a different dynamic bias to the target speaker.
% U 的尺寸不仅会变小，同时它的每一列还 summary 了说话人的信息，且这些列向量对目标说话人有不同的bias

Next, as shown in Fig.\ref{fig:asa} (b), we take
$\mathbf{U}$ and the target speaker embedding vector
$\mathbf{e}^s$ as inputs of the attention module.
The specific procedure is as follows:
\begin{align}
\begin{split}\label{eq:1}
\mathbf{d} ={}&(\mathbf{e}^s)^\mathbf{T} * \mathbf{U}
\end{split}\\
\begin{split}\label{eq:2}
    w_t ={}&\frac{e^{d_t}}{\sum_{i=1}^{T_m}{e^{d_i}}}
\end{split}\\
    \mathbf{B} ={}& \mathbf{e}^s * \mathbf{w}  \label{eq:3}
\end{align}
where $*$ is the operation of matrix multiplication operation,
$\mathbf{T}$ is the operation of transpose,
$\mathbf{d} \in \mathbb{R}^{1 \times T_m}$ is a similarity vector
that measures the correlation between $\mathbf{e}^s$ and $\mathbf{u}_t$, $d_t$ is the $t$-th element of $\mathbf{d}$,
$w_t$ is the softmax of $d_t$ over $t \in [1, T_m]$,
$\mathbf{w} \in \mathbb{R}^{1 \times T_m}$ is the corresponding softmax vector,
and $\mathbf{B} \in \mathbb{R}^{N \times T_m}$ is the attention-based
target speaker embedding matrix.

Each vector at frame $t$ in $\mathbf{B}$ can be regarded as a target speaker-dependent
dynamic bias $\mathbf{b}_t$, because in which the $\mathbf{e}^s$ is weighted by the similarity
$w_t$ that exploits the dynamic target speaker interaction
between the mixture and the adaptation utterance.
In addition, as the dynamic bias $\mathbf{b}_t$ will be very small values
if the mixture speech is occupied by the interferer, this bias may fail
to be a good target speaker guider to supervise the whole network
training.

In order to dynamically adjust the target speaker bias with discriminative
speaker information, as shown in Fig.\ref{fig:asa} (b),
we repeat the target speaker embedding vector $\mathbf{e}^s$ and add
it to the dynamic bias $\mathbf{b}_t$ to get the output $\mathbf{O}$, i.e.,
\begin{equation}	
\begin{split}
\mathbf{o}_t &= \mathbf{b}_t + \mathbf{e}^s\\
\mathbf{O} &= [\mathbf{o}_1, \mathbf{o}_2,..., \mathbf{o}_{T_m}]
\end{split}
\label{eq:add}
\end{equation}
where $\mathbf{O} \in \mathbb{R}^{N \times T_m}$,
$\mathbf{o}_t \in \mathbb{R}^{N \times 1}$ is the $t$-th vector of $\mathbf{O}$ at frame $t$.
This repeat operation in (\ref{eq:add})
assures the target speaker bias won't be too weak, while the
attention mechanism provides a dynamic and more discriminative bias.
Combining these two techniques, the target speaker embedding $\mathbf{O}$
can drive the network towards extracting the target speech more efficiently.

Next, we use the nearest upsampling algorithm to map the $\mathbf{O}$ to the same dimension as the original mixture embedding matrix $\mathbf{Y}$ to get the final target speaker embedding matrix
$\mathbf{E} \in \mathbb{R}^{N \times T}$.
%i.e.,
%\begin{equation}	
%\begin{split}
%\mathbf{o}_i &= \mathbf{e}_{j+(i-1)M}\\
%\mathbf{E} &= [\mathbf{e}_1, \mathbf{e}_2,..., \mathbf{e}_{T}]
%\end{split}
%\end{equation}
%where $i \in [1, T_m]$, $j \in [1, M]$, $\mathbf{e}_t \in \mathbb{R}^{N \times 1}$ is the $t$-th frame vector of $\mathbf{E}$.
Due to the nearest upsampling, the consecutive $M$ vectors (frames) of $\mathbf{E}$ have the same target speaker characteristics, which is consistent with our intuition, namely, the speaker characteristics won't change in a very short duration.

Finally, an element-wise multiplication
between $\mathbf{Y}$ and $\mathbf{E}$ is used to adapt
the network towards the target speaker, i.e.,
\begin{equation}
\mathbf{Y}^s = \mathbf{Y} \odot \mathbf{E}
\end{equation}
where $\mathbf{Y}^s \in \mathbb{R}^{N \times T}$ is the output of the proposed attention-based scaling adaptation.
It is worth noting that, unlike the traditional self-attention that needs
additional learnable weights to project the key, query and value
into high-level representation spaces, our ASA is very simple, and
it is directly performed on the $\mathbf{U}$ and $\mathbf{e}^s$, no 
additional parameter is introduced.

\section{Experiments and results}
\label{sec:exp_res}

\subsection{Dataset}
\label{ssec:dataset}

Our experiments are performed on the simulated spatialized WSJ0 2-mix
corpus \cite{dataset}. All recordings are generated by convolving clean speech
signals with room impulse responses simulated with the image method
for reverberation time ranges from 0.2 s to 0.6 s \cite{td_speakerbeam}.
We use the same way as in \cite{xu} to generate adaptation
utterances of the target speaker. The adaptation utterance is
selected randomly that to be different from the utterance in the mixture.
The adaptation recordings used in our experiments are anechoic.
The size of the training, validation, and test sets
are 20k, 5k, and 3k utterances, respectively.
All of the data are resampled to 8kHz from the original
16kHz sampling rate.

\subsection{Configurations}
\label{ssec:config}

Our experiments are implemented based on the open source
software \cite{funcwj}. We use the same
hyper-parameters as the baseline TD-SpeakerBeam in \cite{td_speakerbeam}.
We set $M=20$ in the ASA process.
All experiments are performed using the SiSDR loss
only ($\alpha=0$ in Eq.(\ref{eq:mtl}))
except when we mention the use of the multi-task loss (MTL),
in which case we set $\alpha=0.5$ to balance the loss tradeoff between the SiSDR and cross-entropy.
For those experiments with IPD combination, the IPD features are extracted
using an STFT window of 32 msec and hop size of 16 msec.
All experiments are performed on the single-channel (first channel)
recordings, except for the experiment with IPD features that extract
from two-channel recordings.
For the performance evaluation, both the
signal-to-distortion ratio (SDR) of BSSeval \cite{sdr} and
the SiSDR \cite{sisdr} are used.

\subsection{Results and discussion}
\label{ssec:results}

\subsubsection{Baseline}
\label{baseline}

We take the single-channel TD-SpeakerBeam \cite{td_speakerbeam}
(without IPD combination) as our baseline.
Results are shown in Table \ref{tab:base}.
System (1) and (2) are the results given in \cite{td_speakerbeam}.
As the source code of
TD-SpeakerBeam is not open resource, we implemented it by ourselves
and reproduced the results as shown in system (3) and
(4) on the same WSJ0 2-mix corpus. We see
that our reproduced results are slightly better than the ones in \cite{td_speakerbeam}.
Moreover, by comparing (1) and (2), or
(3) and (4), there is a big performance gap
between the single-channel and two-channel
TSE tasks, the IPD features extracted from
two-channel recordings are effective to improve the TSE performance.
In addition, we find that the TSE on the same gender conditions
are much more challenge than the mix-gender condition.

\begin{table}[t]
  \caption{SDR / SiSDR (dB) performance of TD-SpeakerBeam (TSB).
   ``IPD" represents the system with internal combination of IPD spatial features,
  ``F" represents female, ``M" represents male, and ``Avg" represents the average results.
	Bold-fonts indicate the best performance.}
  \label{tab:base}

	\footnotesize
	\setlength{\tabcolsep}{0.3mm}
  \centering
  \begin{tabular}{l|c|c|c|c|c}
    \toprule
	System  & IPD & FF & MM & FM & Avg\\
    \midrule
	(1) TSB [14]      & -    & 9.13 / - & 9.47 / - & 12.77 / - & 11.17 / - \\
	(2)             & \checkmark &{\bf 10.17 /-} & 10.30 / - & 12.49 / - & 11.45 / - \\
    \midrule
	(3) TSB (our)       & -    & 9.43 / 8.84 & 10.02 / 9.52 & 12.54 / 12.06 & 11.26 / 10.76\\
	(4)		           & \checkmark & 10.01 / 9.46 &\bf 10.51 / 10.02 & \bf 12.80 / 12.31  & \bf 11.65 / 11.15 \\

  	\bottomrule
  \end{tabular}
\end{table}

\subsubsection{Results of ASA with SiSDR loss}
\label{sisdr_loss}

Table \ref{tab:sisdr_loss} shows the experimental results of
our proposed ASA with SiSDR loss as the system training criterion.
System (1) is our baseline.
In order to examine the effects of the proposed mean pooling
operation that used in the ASA, we remove the blocks of mean
pooling (M) and nearest upsampling (N) used in Fig.\ref{fig:asa} (b).
The proposed attention mechanism is performed directly
between the mixture embedding matrix $\mathbf{Y}$ and
the target speaker embedding vector $\mathbf{e}^s$.
The results are shown in system (2).
Compared with the baseline (system (1)), it's clear to see that system (2)
achieves better performance under all gender-mixing conditions.
The results of system (2) indicate that the dynamic interaction
between the mixture and the target speaker is
helpful to produce a more discriminative speaker bias to
guide the network to extract the target speech.

\begin{table}[h]
  \caption{SDR / SiSDR (dB) performance of
  the proposed attenion based scaling adaptation (ASA) method. ``MP'' represents the mean pooling in ASA.}
  \label{tab:sisdr_loss}
  \footnotesize
	\setlength{\tabcolsep}{0.5mm}
  \centering
  \begin{tabular}{l|c|c|c|c}
    \toprule
	System   & FF & MM & FM & Avg\\
    \midrule
	(1) TSB          & 9.43 / 8.84 & 10.02 / 9.52 & 12.54 / 12.06 & 11.26 / 10.76\\
    \midrule
	(2) ASA            & 9.78 / 9.23& 10.36 / 9.86 & 12.78 / 12.29 & 11.55 / 11.05 \\
	(3) ASA (MP)		           & 9.83 / 9.26  & 10.47 / 9.97 & 12.89 / 12.40 & 11.65 / 11.15 \\
 \midrule
	(4) Para \cite{cd} & 10.83 / 10.21 & 11.52 / 10.99 & 13.12 / 12.61 & 12.26 / 11.72 \\
   (5) Para (ASA) & 11.50 / 10.87 & 12.01 / 11.48 & 13.56 / 13.06 & 12.75 / 12.22 \\
  	\bottomrule
  \end{tabular}
\end{table}

Moreover, as shown in the system (3), the TSE performance
can be further improved by introducing the mean pooling
operation mentioned in Section \ref{sec:asa}.
Summarizing the speaker information
prior to the attention calculation can enhance the ability
of exploiting the dynamic target speaker-dependent information
between different input mixtures and the given adaptation
utterance of the target speaker. Therefore, the mean pooling is set to the default configuration for the rest of our experiments.
Compared with the baseline (system (1)), the proposed ASA (system(3))
achieves $4.2 / 4.7\%$, $4.5 / 4.7\%$ and $2.8 / 1.9 \%$ relative improvements in SDR / SiSDR for female-female, male-male, female-male conditions respectively. It indicates that the ASA can effectively enhance the target speech extraction, especially for the same-gender mixtures.
Furthermore, by comparing the results of system (4) in Table
\ref{tab:base} and system (3) in Table \ref{tab:sisdr_loss},
the proposed system with ASA for single-channel TSE task
achieves competitive performance as the TD-SpeakerBeam
with IPD for the multi-channel task. In addition, please note that the ASA does not introduce any additional learnable parameters like linear transformation weight of query, key, and value in self-attention \cite{self-att}. The proposed ASA only contains pure mathematical operations.
These promising results indicate that our proposed ASA is effective to improve
the discrimination of target speaker clues for TSE tasks.

Although as suggested in \cite{ms}, the local dynamics and
the temporal structure will be lost due to the
averaging operation, the averaging can provide a more
robust speaker bias vector than a single frame
vector. Experimental results show that the local mean
pooling and the nearest upsampling operations are
beneficial for ASA to exploit the target speaker clues
in a more efficient way. We also tried applying the attention mechanism from \cite{ms} to the TSB structure, but the training process had difficulties to converge and so we stopped it after several epochs.

Furthermore, we also investigate effects of ASA on two-channel parallel encoder based TSB system \cite{cd, para} which directly sums the waveform encodings of each input channel to enhance the mixture representation. The adaptation methods of system (4) and (5) are SA and ASA with mean pooling. Results show that the extraction performance can be further improved by introducing the proposed ASA mechanism.

\subsubsection{Results of ASA with multi-task loss}
\label{mtl_loss}

\begin{table}[t]
  \caption{SDR / SiSDR (dB) performance of
  the proposed attention based scaling adaptation (ASA) method with
  multi-task loss (MTL).}
  \label{tab:mtl_loss}
  \footnotesize
	\setlength{\tabcolsep}{0.5mm}
  \centering
  \begin{tabular}{l|c|c|c|c}
    \toprule
	System  & FF & MM & FM & Avg\\
    \midrule
	(1) TSB       & 9.43 / 8.84 & 10.02 / 9.52 & 12.54 / 12.06 & 11.26 / 10.76\\
	(2) TSB (MTL) & 9.66 / 9.09 & 10.33 / 9.84 & 12.75 / 12.26 & 11.51 / 11.00 \\
    \midrule
	(3) ASA (MTL)           & \bf 9.93 / 9.35 & \bf 10.63 / 10.13 & \bf 12.92 / 12.43& \bf 11.73 / 11.22 \\
  	\bottomrule
  \end{tabular}
\end{table}

We also investigated the performance of ASA with multi-task loss.
Results are shown in Table \ref{tab:mtl_loss}.
System (1) is the TD-SpeakerBeam trained with SiSDR loss function,
and system (2) and (3) are the TD-SpeakerBeam and the proposed ASA system (with mean pooling)
trained with a multi-task loss respectively.
Results show that even using a multi-task loss, the proposed ASA
still achieves better performances than the baselines.
Note that in this case, we scaled the attention-based target speaker
embedding matrix $\mathbf{B}$ by $\sqrt{N}$ as we
found this combination can result in better performance.

\section{Conclusion}
\label{sec:conclusion}

In this work, we propose a novel target speaker adaptation
technique for time-domain target speech extraction tasks.
A special attention mechanism is designed to effectively
exploit the dynamic target speaker-dependent
interaction between different mixtures and the given adaptation
utterance. This dynamic information can improve
the target speaker clues and its discriminations for target
speech extraction. Experiments on the spatialized WSJ0 2-mix
corpus demonstrate that our proposed method
effectively improved the TD-SpeakerBeam for target speech extraction.
Furthermore, it is surprising to find that the single-channel
performance gains result from the proposed ASA
are competitive with the multi-channel TD-SpeakerBeam with IPD features.
Our future work will focus on how to combine the proposed method
with the multi-task loss in a more appropriate way.

\bibliographystyle{IEEEbib}
\bibliography{refs}

\begin{thebibliography}{10}

\bibitem{dpcl}
J.~R. Hershey, Z.~Chen, J.~Le~Roux, and S.~Watanabe,
\newblock ``Deep clustering: Discriminative embeddings for segmentation and
  separation,''
\newblock in {\em 2016 IEEE International Conference on Acoustics, Speech and
  Signal Processing (ICASSP)}. IEEE, 2016, pp. 31--35.

\bibitem{dan}
Z.~Chen, Y.~Luo, and N.~Mesgarani,
\newblock ``Deep attractor network for single-microphone speaker separation,''
\newblock in {\em 2017 IEEE International Conference on Acoustics, Speech and
  Signal Processing (ICASSP)}. IEEE, 2017, pp. 246--250.

\bibitem{pit}
M.~Kolb{\ae}k, D.~Yu, Z.~H. Tan, and J.~Jensen,
\newblock ``Multitalker speech separation with utterance-level permutation
  invariant training of deep recurrent neural networks,''
\newblock {\em IEEE/ACM Transactions on Audio, Speech, and Language
  Processing}, vol. 25, no. 10, pp. 1901--1913, 2017.

\bibitem{tasnet}
Y.~Luo and N.~Mesgarani,
\newblock ``{Conv-TasNet}: Surpassing ideal time--frequency magnitude masking
  for speech separation,''
\newblock {\em IEEE/ACM transactions on audio, speech, and language
  processing}, vol. 27, no. 8, pp. 1256--1266, 2019.

\bibitem{tasnet_relate1}
D.~Ditter and T.~Gerkmann,
\newblock ``A multi-phase gammatone filterbank for speech separation via
  {TasNet},''
\newblock in {\em 2020 IEEE International Conference on Acoustics, Speech and
  Signal Processing (ICASSP)}. IEEE, 2020, pp. 36--40.

\bibitem{tasnet_relate2}
J.~Heitkaemper, D.~Jakobeit, C.~Boeddeker, L.~Drude, and R.~Haeb-Umbach,
\newblock ``Demystifying {TasNet}: A dissecting approach,''
\newblock in {\em 2020 IEEE International Conference on Acoustics, Speech and
  Signal Processing (ICASSP)}. IEEE, 2020, pp. 6359--6363.

\bibitem{tasnet_relate3}
T.~Ochiai, M.~Delcroix, R.~Ikeshita, K.~Kinoshita, T.~Nakatani, and S.~Araki,
\newblock ``{Beam-TasNet}: Time-domain audio separation network meets
  frequency-domain beamformer,''
\newblock in {\em 2020 IEEE International Conference on Acoustics, Speech and
  Signal Processing (ICASSP)}. IEEE, 2020, pp. 6384--6388.

\bibitem{tasnet_relate4}
M.~Pariente, S.~Cornell, A.~Deleforge, and E.~Vincent,
\newblock ``Filterbank design for end-to-end speech separation,''
\newblock in {\em 2020 IEEE International Conference on Acoustics, Speech and
  Signal Processing (ICASSP)}. IEEE, 2020, pp. 6364--6368.

\bibitem{tasnet_relate5}
R.~Gu, S.-X. Zhang, L.~Chen, Y.~Xu, M.~Yu, D.~Su, Y.~Zou, and D.~Yu,
\newblock ``Enhancing end-to-end multi-channel speech separation via spatial
  feature learning,''
\newblock in {\em 2020 IEEE International Conference on Acoustics, Speech and
  Signal Processing (ICASSP)}. IEEE, 2020, pp. 7319--7323.

\bibitem{tasnet_relate6}
S.~Sonning, C.~Sch{\"u}ldt, H.~Erdogan, and S.~Wisdom,
\newblock ``Performance study of a convolutional time-domain audio separation
  network for real-time speech denoising,''
\newblock in {\em 2020 IEEE International Conference on Acoustics, Speech and
  Signal Processing (ICASSP)}. IEEE, 2020, pp. 831--835.

\bibitem{sbfe}
M.~Delcroix, K.~Zmolikova, K.~Kinoshita, A.~Ogawa, and T.~Nakatani,
\newblock ``Single channel target speaker extraction and recognition with
  speaker beam,''
\newblock in {\em 2018 IEEE International Conference on Acoustics, Speech and
  Signal Processing (ICASSP)}. IEEE, 2018, pp. 5554--5558.

\bibitem{spex}
C.~Xu, W.~Rao, E.~S. Chng, and H.~Li,
\newblock ``{SpEx}: Multi-scale time domain speaker extraction network,''
\newblock {\em IEEE/ACM Transactions on Audio, Speech, and Language
  Processing}, vol. PP, no. 99, pp. 1--1, 2020.

\bibitem{td_speakerbeam}
M.~Delcroix, T.~Ochiai, K.~Zmolikova, K.~Kinoshita, N.~Tawara, T.~Nakatani, and
  S.~Araki,
\newblock ``Improving speaker discrimination of target speech extraction with
  time-domain speakerbeam,''
\newblock in {\em 2020 IEEE International Conference on Acoustics, Speech and
  Signal Processing (ICASSP)}. IEEE, 2020, pp. 691--695.

\bibitem{ms}
X.~Xiao, Z.~Chen, T.~Yoshioka, H.~Erdogan, C.~Liu, D.~Dimitriadis, J.~Droppo,
  and Y.~Gong,
\newblock ``Single-channel speech extraction using speaker inventory and
  attention network,''
\newblock in {\em 2019 IEEE International Conference on Acoustics, Speech and
  Signal Processing (ICASSP)}. IEEE, 2019, pp. 86--90.

\bibitem{tencent}
X.~Ji, M.~Yu, C.~Zhang, D.~Su, T.~Yu, X.~Liu, and D.~Yu,
\newblock ``Speaker-aware target speaker enhancement by jointly learning with
  speaker embedding extraction,''
\newblock in {\em 2020 IEEE International Conference on Acoustics, Speech and
  Signal Processing (ICASSP)}. IEEE, 2020, pp. 7294--7298.

\bibitem{self-att}
A.~Vaswani, N.~Shazeer, N.~Parmar, J.~Uszkoreit, L.~Jones, A.~N. Gomez,
  L.~Kaiser, and I.~Polosukhin,
\newblock ``Attention is all you need,''
\newblock {\em arXiv preprint arXiv:1706.03762}, 2017.

\bibitem{ipd}
Z.~Chen, X.~Xiao, T.~Yoshioka, H.~Erdogan, J.~Li, and Y.~Gong,
\newblock ``Multi-channel overlapped speech recognition with location guided
  speech extraction network,''
\newblock in {\em Spoken Language Technology Workshop (SLT)}. IEEE, 2018, pp.
  558--565.

\bibitem{adap}
M.~Delcroix, K.~Zmolikova, T.~Ochiai, K.~Kinoshita, S.~Araki, and T.~Nakatani,
\newblock ``Compact network for speakerbeam target speaker extraction,''
\newblock in {\em 2019 IEEE International Conference on Acoustics, Speech and
  Signal Processing (ICASSP)}. IEEE, 2019, pp. 6965--6969.

\bibitem{sisdr}
J.~Le~Roux, S.~Wisdom, H.~Erdogan, and J.~R. Hershey,
\newblock ``{SDR}--half-baked or well done?,''
\newblock in {\em 2019 IEEE International Conference on Acoustics, Speech and
  Signal Processing (ICASSP)}. IEEE, 2019, pp. 626--630.

\bibitem{dataset}
Z.-Q. Wang, J.~Le~Roux, and J.~R. Hershey,
\newblock ``Multi-channel deep clustering: Discriminative spectral and spatial
  embeddings for speaker-independent speech separation,''
\newblock in {\em 2018 IEEE International Conference on Acoustics, Speech and
  Signal Processing (ICASSP)}. IEEE, 2018, pp. 1--5.

\bibitem{xu}
  \url{https://github.com/xuchenglin28/speaker_extraction/tree/master/simulation}.

\bibitem{funcwj}
 \url{https://github.com/funcwj/conv-tasnet}.

\bibitem{sdr}
E.~Vincent, R.~Gribonval, and C.~F{\'e}votte,
\newblock ``Performance measurement in blind audio source separation,''
\newblock {\em IEEE transactions on audio, speech, and language processing},
  vol. 14, no. 4, pp. 1462--1469, 2006.

\bibitem{cd}
J.~Han, W.~Rao, Y.~Wang, and Y.~Long,
\newblock ``Improving channel decorrelation for multi-channel target speech
  extraction,''
\newblock in {\em Proc. Interspeech}, 2021, pp. 1847--1851.

\bibitem{para}
R.~Gu, J.~Wu, S.-X. Zhang, L.~Chen, Y.~Xu, M.~Yu, D.~Su, Y.~Zou, and D.~Yu,
\newblock ``End-to-end multi-channel speech separation,''
\newblock {\em arXiv preprint arXiv:1905.06286}, 2019.

\end{thebibliography}

\end{document}